\documentclass[graybox]{svmult}

\usepackage{type1cm}        
\usepackage{makeidx}         
\usepackage{graphicx}        
\usepackage{multicol}        
\usepackage[bottom]{footmisc}

\usepackage{newtxtext}       %
\usepackage[varvw]{newtxmath}       

\usepackage[numbers,sort&compress]{natbib}
\usepackage{todonotes}

\makeindex             

\begin{document}

\title*{A simplicity bubble problem and zemblanity in digitally intermediated societies\thanks{ This article is based on, and extends the abstract \cite{Abrahao2023SBZemblanityarxiv} approved in the \emph{Ninth Conference on Model-Based Reasoning, Abductive Cognition, Creativity}, 2023.}}
\author{Felipe S. Abrah\~{a}o, Ricardo P. Cavassane, Michael Winter, Mariana Vitti-Rodrigues, and Itala M. L. D'Ottaviano}
\institute{
Felipe S. Abrah\~{a}o 
\at Oxford Immune Algorithmics, Oxford University Innovation, U.K.,\\ \email{Felipe.Abrahao@algocyte.ai}  
\at DEXL, National Laboratory for Scientific Computing, Brazil.
\at Centre for Logic, Epistemology and the History of Science, University of Campinas, Brazil.
\at Algorithmic Nature Group, LABORES, France. 
\and
Ricardo P. Cavassane, and Itala M. L. D'Ottaviano 
\at Centre for Logic, Epistemology and the History of Science, University of Campinas, Brazil,\\ 
\email{ricardo.peraca@gmail.com}; \email{itala@unicamp.br}
\and 
Michael Winter \at Independent Researcher, Germany,\\ \email{mwinter@unboundedpress.org}
\and 
Mariana Vitti Rodrigues \at São Paulo State University, Brazil,\\ \email{mvittirodrigues@gmail.com}
}

%
%
\maketitle


\abstract{
In this article, we discuss the ubiquity of Big Data and machine learning in society and propose that it evinces the need of further investigation of their fundamental limitations.
We extend the ``too much information tends to behave like very little information'' phenomenon to formal knowledge about lawlike universes and arbitrary collections of computably generated datasets. This gives rise to the simplicity bubble problem, which refers to a learning algorithm equipped with a formal theory that can be deceived by a dataset to find a locally optimal model which it deems to be the global one. 
In the context of lawlike (computable) universes and formal learning systems, we show that there is a ceiling above which formal knowledge cannot further decrease the probability of zemblanitous findings,
should the randomly generated data made available to the formal learning system be sufficiently large in comparison to their joint complexity.
Zemblanity, the opposite of serendipity, is defined by an undesirable but expected finding that reveals an underlying problem or negative consequence in a given model or theory, which is in principle predictable in case the formal theory contains sufficient information.
We also argue that this is an epistemological limitation that may generate unpredictable problems in digitally intermediated societies.
}


%
\section{Introduction}
\label{labelIntro}

The advance of technologies for collecting, mining, storing, retrieving, and programmatically analyzing massive amounts of data has enabled an increasing automation of scientific practices, bringing the promise of facilitating the discovery of patterns in huge datasets. 
In recent years, the focus on data science has moved toward artificial intelligence (AI), specifically ``bottom-up'' approaches in machine learning such as generative AI (GenAI) and large language models (LLMs).
For example, the deep learning algorithm called \emph{AlphaFold2}, whose function is to predict possible protein structures in three dimensions from amino acid sequences in one dimension, has enabled discoveries in several areas such as biochemistry, pharmacology, biomedicine, among others \cite{Callaway2022AlphaFoldAI}. These advances make scientists confront new challenges in scientific methodologies and everyday practices in order to properly benefit from the above technologies \cite{Zenil2023AIReviewarxivv3}. For example, in the  case of AlphaFold2, it is now up to the researcher to determine the negative and positive potential of these newly found protein structures without fully understanding how they function.

The large amount of collected data coupled with the success of AI algorithms is rapidly increasing the pervasiveness of computational intermediation of humans and institutional relationships.
Some enthusiasts of the use of new technologies for data collection and data analysis argue that the access to ever-increasing volumes of diverse data has made the formulation of explanatory hypotheses and the development of scientific theories, along with the elaboration of causal models, obsolete \cite{Anderson2008Endoftheory}. 
In this sense, Big Data enthusiasts imply and believe that the data analysis is capable of making serendipitous discoveries \cite{Cavassane2022}: findings that are unpredictable, random, or appear to be made out of the blue with respect to the data analyst's scientific skills or current state of knowledge, but nevertheless happen to be fortunate in the end. 
Serendipity combines both being an accident and the recognition of such accident as a valuable source for knowledge acquisition~\cite{Merton2000SocialTheorySocial}.
Being fortuitous, serendipity would enable one to solve a previously stated problem in an unforeseen manner or develop irreducibly new theoretical findings \cite{Cavassane2022}.


However, in the context of Big Data analysis, we argue that a zemblanatious finding should be more prominent than a serendipitous one.
Zemblanity is the polar opposite of serendipity \cite{Boyd1998}. While serendipity refers to an unexpected or unpredictable event that presents an opportunity for a fortunate discovery, zemblanity refers to a result that should have been expected/predictable which reveals an underlying problem or negative consequence in a given model or theory. It is not a discovery at all, but rather a systemic flaw or negative aspect that the observer is either intentionally or unintentionally overlooking, not acknowledging in the results' statements, or somehow refusing to understand. This is because, both in principle and in practice, a zemblanitous phenomenon would be predictable by the data analyst, being one of the possible direct implications of (or one that can be inferred from) the current theory and state of knowledge. 
Unfortunately, these deleterious or problematic properties are often only revealed when these conclusions are put into practice in society and cause societal conflict and damage. 

We base this finding on 
the paradox-like Big Data problem in which
``too much information tends to behave like very little information'' \cite{Smith2020,Calude2017} and believe that it is a novel fundamental limitation of formal learning systems \cite{Abrahao2021dSBpaperarxiv}. The Big Data ``paradox'' \cite{Smith2020} refers to the phenomenon in which
although any formal theory, learning algorithm, and/or statistical method would benefit from large amounts of accurate and fine-grained data points, the larger the dataset the more likely it is to discover/detect spurious patterns or correlations \cite{Calude2017}. 
A spurious correlation is one that would have occurred (or is expected to occur) in randomly generated data anyway. It cannot provide useful information about the distinctive characteristics of the dataset or process that generates it. 
Such findings may deceive or mislead an observer into thinking a statistically significant pattern was discovered. This particular paradox-like problem is an example of a zemblanitous event: an undesirable but expected finding \cite{Cavassane2022}.

On the one hand, serendipitous findings occur by chance, or are unpredictable, but are valuable with respect to solving a previously stated problem by introducing new and promising hypotheses or lines of investigation that can be used to develop a new theory or extend previous theories \cite{Merton1948}. 
On the other hand, a zemblanitous finding is a result of biases or errors that will necessarily appear in the model. Thus, since the term ``zemblanity" means the complementary antonym of ``serendipity", a zemblanitous finding is not the product of chance, but actually inherent to these Big Data technologies.\footnote{A second type of zemblanity could consist not in the complementary antonym of “serendipity”, but in its gradative antonym, that is, meaning not a necessary result, but one which, though expected to eventually happen, is also triggered by an unexpected event. One could thus admit intermediary degrees between both notions, with an unexpected and valuable finding generated by an accident with potentially positive outcomes (since serendipity can happen by accident and sagacity~\cite{WalpoleCorrespondence,Andel1994AnatomyUnsoughtFinding}, or at the intersection of chance and wisdom~\cite{Copeland2019}) on one side of the scale; and an expected and negative result generated by an accident with potentially negative outcomes on the other side of the scale. Further investigation on this alternative definition is a necessary future work.}

Big Data analysis is prone to zemblanitous events because our current mathematical theories are capable of proving that one can find patterns---and consider them statistically significant---that are already expected to occur for underlying data generating processes that are random.
If the observer or scientist is not a priori aware of the spurious patterns that may emerge in random scenarios, then statistical machine learning methods, even when fed with large amounts of data such as in the Big Data paradigm, are prone to zemblanity instead of being a new and enhanced source of serendipity.
This differs from the claim that, just by ``letting the data speak",  the analysis of Big Data would easily produce unexpected and useful insights \cite{Cukier2013} and foster serendipitous discoveries.

The phenomenon of ``too much information tends to behave like very little information''
is in consonance with on aspect of machine learning that developers of the technology have grappled with since the inception of AI: avoiding overfitting and guaranteeing that the learner will not get stuck in local optima. 
In this article, we aim to show that these issues are the result of a systemic problem: that there is a deceiving phenomenon inherent to these systems that results from a fundamental limit of formal knowledge systems.

Overfitting refers to the case in which the error on the dataset against which the trained model candidate is tested is considered to be relatively higher than the training error, which was employed (and previously chosen) as the error bounds or margins in the training phase so the learner would classify a model candidate as successful with regards to the desired task.
On the other hand, underfitting occurs when the value of the loss function (i.e., the function that calculates how much a model candidate diverges from the optimal one, or from the target values) retrieved from the training dataset is considered to be higher than that training error.
Underfitting can be easily solved by increasing the number, or the complexity, of the parameters, hyperparameters, layers of nodes in a neural network, etc. Overfitting demands much more sophisticated methods along with human supervision and agency. 
These methods must effect a procedure, such as regularization, that adapts or translates some heuristics, inspired by Occam's razor or a bias toward simplicity, into a successful and tractable computational implementation of an algorithm that performs the desired task within the required error bounds \cite{Goodfellow2016,Witten2017}.
By embedding such a bias toward simplicity into the regularization phase itself, recent results have shown that models yielding a perfect fit with a minimum training error for every data point in the training set tend to naturally have greater algorithmic complexity so that
overfitting can be avoided by design \cite{HernandezOrozco2021}.

The no free lunch theorem \cite{Wolpert1997} says that no learning algorithm is any better than any
other equally likely learning algorithm \cite{Goodfellow2016}.
When averaging out all possible optimization problems, this means that for every learner there will exist a task at which the learner succeeds as much as another learner that already fails at it.
This implies a theoretical limitation where one cannot guarantee that there is a general method in which the learner avoids getting stuck in locally optimal models.
That is, one cannot guarantee that there is a general method in which the learner can always approximate the globally optimal solution within the desired error margins.
In practice, one can assume distributions in real-world data gathering that favor simpler models, thus avoiding the condition of performing the uniform average.
This bias toward simplicity constitutes a way out of the conditions for the no free lunch theorem \cite{Goodfellow2016,Witten2017,Scholkopf2021} because in practice one can reduce the scope of the problem by imposing a priori constraints on the possible model candidates.

The local \emph{versus} global optima problem remains a challenge that machine learning methods must overcome \cite{Goodfellow2016}, both in theory and empirically, which raises the necessity of explaining why this problem is still occurring \cite{Abrahao2021dSBpaperarxiv}.
In order to understand how the overfitting problem and the local \emph{versus} global optima problem intersect, we tackle the question of whether or not the bias toward simplicity, or a nomic assumption alone, can guarantee that the prediction made by the learner will continue to be non over-fitted not only for the available phenomena, but also for other phenomena to which the learning process has no access prior to the estimation of the optimal model \cite{Abrahao2021dSBpaperarxiv,Arroyo2023Nomicandsimplicity}.
This is also where a scientific and technological problem reflects an epistemological problem regarding the theoretical limitations of computationally assisted learning processes, whether guided or not by an arbitrarily chosen formal theory.
The results in \cite{Abrahao2021dSBpaperarxiv} demonstrate that there is a simplicity bubble into which lawlike universes can trap the observer (in this case, the formal learning system) into being deceived to find an optimal solution that it deems globally optimal although never being able to actually approximate the globally optimal solution.
A formal learning system is defined by an arbitrary machine learning algorithm equipped with an arbitrary formal theory that has access to an external source of events (encoded into datasets) for which it can produce a model or theory as an optimum candidate \cite{Abrahao2021dSBpaperarxiv}.

As we will discuss in the following, this reveals itself as an extension of the
``too much information tends to behave like very little information'' phenomenon in sufficiently large datasets
that implies the existence of a ceiling above which formal systems with access to external universes (or sources of events) cannot increase the degree of validity from corroborating outcomes of hypothesis testing.
Formal learning systems cannot in general guarantee (with probability as low as one wishes) that their optimal solutions, models, or theories are considered to be optimal because they actually better predict future or yet unavailable phenomena (i.e., those that differ from the phenomena encoded into the datasets from which the formal learning system produced the optimal model candidate).  
On the contrary, formal learning systems may classify a model candidate as optimal while at the same time as arbitrarily diverging from optimally predicting yet unknown phenomena, that is, diverging from the actual globally optimal solution for the external source.

Such limitations are systemic to any formal-theoretic or computational method that learns through experience in arbitrarily large lawlike universes.
By experience\footnote{Note that in the ``metaphenomenological'' model proposed in \cite{Winter2020}, the concept of the digital experience is originally defined in reverse: it is an algorithm that takes as input a subject (i.e., in our present case, a model) and always transforms it into something else. This theoretically would allow the subject to overcome the problem of getting trapped in local optima, but only with a limited use of an oracle to avoid non-halting results similar to Chaitin's metabiological model~\cite{Chaitin2012}. The definition employed in this article so as to more accurately reflect the way machine learning algorithms currently function.}, we mean a set of events input to a learning algorithm equipped with a formal theory \cite{Winter2020}.
A lawlike universe follows a nomic assumption about its own structure: that it can be explained or generated by a finite set of generative rules that are much simpler than the whole set of events or outcomes~\cite{Arroyo2023Nomicandsimplicity}. 
In particular, this is the case of any possible universe that is in principle computable \cite{Zenil2013,Abrahao2021bpublished}---which holds even if one might never know the universe's original generative algorithm. 

We also argue in this article that implications of such theoretical limitations extend to societal impacts and advocate that more scientific and philosophical research of these findings is paramount in order to tackle an emerging concern in digitally intermediated societies.

\section{The simplicity bubble problem}
\label{sectionSimplicitybubble}

In general terms, a model is considered to be optimal if its error rate, as computed by some performance measure, is minimized on a collection of data points.
A model is locally optimal if it is considered to be optimal for a constrained number of data points (i.e., a proper subset of an entire universe); and globally optimal if it is optimal for all possible data points that can be generated by the external source (i.e., the entire universe; past, present to future) \cite{Goodfellow2016}.
A deceiving dataset made available to a learning algorithm is one from which the learning algorithm (whether equipped or not with an arbitrarily chosen formal theory) \emph{cannot} find a model that performs optimally on fresh data, although it does find one that is considered to perform optimally on the available dataset \cite{Abrahao2021dSBpaperarxiv}.
That is, the learning algorithm is indeed able to generate an optimal model, and it generates one that completely satisfies the previously chosen performance criteria to be considered a global one.
However, this model is in fact a locally optimal model.
In this context, if the actual globally optimal model 
is also unpredictable (given the available data and a learning algorithm equipped with an arbitrarily chosen formal theory), then we call the corresponding deceiving dataset an \emph{unpredictable deceiver}.  

For a sufficiently large dataset given an arbitrarily chosen learning algorithm and a formal theory, it has been shown that the probability of a dataset being an unpredictable deceiver dominates (except for an independent multiplicative constant) the probability of any other randomly generated dataset of equal or greater size~\cite{Abrahao2021dSBpaperarxiv}.
This demonstrates that in the context of randomly generated lawlike universes, no learning algorithm and formal theory can make the probability of unpredictably deceiving datasets smaller than a certain constant, which only depends on the learning algorithm and formal theory.

The condition for this to hold (i.e., the one that imposes a lower bound on the size of the datasets) guarantees that the algorithmic information content \cite{Abrahao2021bpublished} of any other dataset containing unavailable or unknown phenomena in the form of data points will be sufficiently high so that it can render the original dataset a deceiver.
And this will be the respective deceiver that induces the learning algorithm to find a locally optimal model whose algorithmic information content is relatively much smaller. 
The underlying main idea here is to construct a sufficiently large dataset, available to the learning algorithm, such that the prediction error (i.e., the overall generalization error taking into account fresh data) cannot be minimized beyond a certain level by the learning algorithm which was already successfully trained by the available data.

As a result, large enough datasets can deceive learning algorithms into accurately finding a (locally) optimal solution (which the learning algorithm may return as being a global optimum candidate) while never being able to actually find a globally optimal solution for the whole universe of possible events. 
This is a \emph{simplicity bubble effect} (SBE) \cite{Abrahao2021dSBpaperarxiv}: 
the phenomenon where the underlying data generating process (mechanism or external source) is trapping an algorithm (or a formal theory) into a \emph{simplicity bubble} because the data made available by the generating process is complex enough to ensure that any model proposed by the algorithm (or formal theory) is insufficiently complex to successfully approximate a globally optimal model that actually corresponds to any fresh data that the algorithm may receive. 
The SBE has to do with a mathematical phenomenon that appears from a (algorithmic) complexity-centered interplay between data and algorithms.
Another form of interpreting this idea is that the deception originates from a wide enough gap between the algorithmic information of the underlying generative model, laws, or principles of the external source of the events or phenomena and the algorithmic information of the formal learning systems, i.e., the observer.
This gap is guaranteed when the available dataset is sufficiently large.
As more and more data becomes available, its underlying generating process (or mechanism) likely becomes more complex than the observer's tools or formal knowledge themselves \cite{Abrahao2021dSBpaperarxiv}.
Only by adding a larger amount of (unpredictable) fresh data, i.e., new phenomena, not yet available to the observer in the first place would the observer hypothetically be capable of ``bursting'' this bubble. Only then, would the observer ``realize'' that the global optimum proposed by the model was not global at all.

\subsection{No formal learning system can rule out demons in justified theories for lawlike universes}\label{sectionDemons}

The SBE can be extended to a general epistemological limitation of knowledge production systems, should one assume that: 

\begin{description}[Condition I]
\item[Condition I]{the external phenomena to the observers occur in lawlike (computable) universes in which the observers do not a priori know, or do not have ultimate access to, the original and actual generative algorithm, laws, or theory of everything;}

\medskip

\item[Condition II]{observers' predictions and explanations (in the form of an optimal model candidate or an optimal theory candidate) along with their respective reliability, validity, or other justification criteria can only be asserted within the scope of formal theories and computational methods.}
\end{description}

This epistemological limitation constitutes a ceiling (for the amount of events or phenomena that can corroborate a theory or model) above which one cannot further guarantee its theories have sufficient formal knowledge about the yet unknown phenomena in lawlike universes \cite{Arroyo2023Nomicandsimplicity}.
As traditionally assumed in the justification of induction problems as one of the limitations of current scientific methods, the sheer fact of the known set of phenomena being explained and thus corroborating a formal theory's predictions does not guarantee its ultimate validity.
Scientific practice and common sense in the respective area of knowledge would lead one to increase the confidence level or potential validity of a previous theory as long as a larger number of distinct events corroborate and helps justify it.
According to Condition II, in the context of formal learning systems and lawlike universes, the SBE counterintuitively shows this cannot continue beyond a certain sufficient large amount of known data.
This holds because any scientific statement on the justification of the optimal theory candidate can only be formalized by formal theories and computational methods. 

Once one hits the aforementioned ceiling, one cannot keep increasing the confidence level or potential validity of a previous theory any longer.
This may seem paradoxical at first glance because lawlike universes should be those about which more corroboration of previous theories should lead to more confidence that one is indeed approaching the underlying yet unknown laws that governs the behavior of the events or phenomena in the universe.
What the existence of the SBE shows is that this does not apply in general when the amount of already known and successfully explainable phenomena (i.e., the available dataset to the formal learning system from which it achieved the optimal model or theory candidate) gets sufficiently large in comparison to the complexity of the current formal knowledge and computational methods. 

Let a hypothetical abstract entity, i.e., a `demon', in a thought experiment correspond to a sufficiently large dataset (or to the underlying generative and governing agent for the set of every possible phenomena) which is an unpredictable deceiver that triggers the SBE for a given observer, i.e., a formal learning system.
The above epistemological limitation is essentially a lower bound for the probability (given any arbitrary formal theory that defines the arbitrarily chosen probability measure) that such a demon exists.
As the set of known and explainable phenomena becomes arbitrarily large, if Conditions I and II are satisfied, no one can rule out the possibility of those demons by endlessly decreasing the probability of their existence.

\section{Interplay between Big Data, formal learning systems, and humanity}\label{sectionComputersinsociety}

Beyond the context of theoretical computer science or hypothetical limits of computation-assisted scientific methods, we discuss in this section potential implications in digitally intermediated societies. To do so, we consider a social system dynamics for which the yet unknown novel phenomena, which are unpredictable because of the SBE, exhibit emergent behavior \cite{Abrahao2021bpublished,Abrahao2020bsimplelocalrules}.

Within the same context of the results we have discussed in Section~\ref{sectionSimplicitybubble}, it is demonstrated in \cite{Abrahao2021bpublished} that the emergence of algorithmic information is the strongest form of emergence that formal theories can explain.
In addition, both the synchronic and the diachronic variants of emergent phenomena were demonstrated to occur in networked systems and evolutionary systems, respectively. 
The emergence of novel algorithmic information content refers to new knowledge that formal theories irreducibly require in order to explain systemic behavior
\cite{Abrahao2021bpublished}.
Those emergent phenomena in real-world complex systems, such as societies whose policies are guided by AI and the massive gathering of personal data, may have unexpected implications \cite{Cavassane2023BigDataSelffulfillingproph}.

Contrary to the idea that data are objective instances of the object they represent, Leonelli \cite{Leonelli2014QuantityepistemologyBigData,Leonelli2023OpenScience} stresses that the epistemological value of data depends on the context of their origin, the objective that guided its collection, as well as the potential use and reuse of data in different contexts. 
The process of data standardization that is conducted in order to apply computational techniques involves choices that are hardly possible to automate once it depends on the social-cultural, as well as local knowledge and research environments in which the data is collected, mined, and curated. 
This is exemplified in fundamental physical problems such as the correct decoding of technosignatures of other intelligent agents for which we do not have a priori knowledge about their computational methods and mathematical formalisms \cite{Zenil2023arxivETpaper}. If certain conditions are not met, such changes in their computational methods may be so dramatic that their signal would be interpreted as noise.

Leonelli proposes the notion of data journeys \cite{Leonelli2014QuantityepistemologyBigData,Leonelli2023OpenScience} to account for the material aspects, local idiosyncrasies, experimental settings, inferential processes, among other elements which influence the results in data-intensive research related to the processes of data achievement, integration, and use.
Data journeys involve the process of de-contextualization of data from their local origin, such that it becomes digital and integrated in structured or semi-structured datasets that can be retrieved in other domains, allowing the possibilities of data reuse, reanalysis, recombination, repurposing, and repositioning \cite{Leonelli2014QuantityepistemologyBigData,Collmann2016EthicalBigDataaouthumans}. 

An example of a data journey is the process of retrieval of information about amino acid sequences stored in genomic data infrastructures in order to design new forms of protein structures using freely available software for academic research \cite{Anishchenko2021denovodesigndeeplearning}. The storage of protein data in massive datasets, such as \emph{Uniprot}, requires the process of standardization of the uploaded forms of proteins, being experimentally achieved or computationally designed. In addition, the dataset can impose some limitation in the retrieval of information. For example, a maximum of one hundred sequences per search. In the endeavor to cover as many species as possible, there could also be biases which favor the species that are well studied over fewer known organisms.
Other kinds of biases may arise in the analysis phase itself as a limitation of statistics-based methods as is demonstrated in \cite{Uthamacumaran2023arxivMSpaper} when one tries to distinguish living from non-living matter.

As presented in \cite{Cavassane2023BigDataSelffulfillingproph}, a potential problem illuminated by data journeys is the growing reliance on acknowledgeable datasets for decision-making processes in a software model which might be used to inform further research whose results might, in a feedback loop, serve as input to the data infrastructure. Our idea here is not to disregard the epistemic value of data integration, data modelling, and data retrieval, but to shed light on possible challenges of the growing automation of scientific practices by reflecting upon decision-making processes involved in data journeys.
These challenges and a proposal on how to tackle them are discussed in \cite{Zenil2023AIReviewarxivv3}, in which new theoretical and technological tools are proposed in order to deal with the dynamics between AI-driven scientific discoveries and human scientists themselves once both will belong to the same knowledge production system in a closed-loop manner.
We need to provide ways for making a given system accountable by its results in order to guide decision-making processes which can directly or indirectly affect society~\cite{Alvarado2023SimulatingScienceComputer}.

In summary, data journeys constitute dynamic structures that, in association with computational tools, might guide scientific research and decision-making processes, promising new forms of scientific discovery. Once suitable for data journeys, data can grow in meaning by the process of recontextualization, with their original purpose being revised to fit a different research question or to redirect the course of investigation. 
Similar to the framework given by algorithmic information dynamics (AID) \cite{Zenil2020cAIDscholarpedia,Abrahao2021bpublished,Zenil2019dAIDcalculusiScience,Zenil2018,Zenil2019} which studies causal inference without the a priori separation between observers and objects, data journeys contrast the so-called ‘objective-oriented’ view of traditional scientific practice. One in which, by avoiding human interference and potential mistakes in scientific practice, we would eventually achieve neutral scientific results. This argument is succinctly summarized by Leonelli: ``the paradox consists of the observation that, despite their epistemic value as `given', data are clearly made'' \cite{Leonelli2014QuantityepistemologyBigData}.
Instead of this paradox suggesting an inexorable limit that science should acknowledge, novel methodologies should embed this paradigm shift in order to allow the investigation of both the objects and the observers themselves as interacting systems from the beginning and as part of the same theoretical and empirical framework \cite{Zenil2020cAIDscholarpedia,Abrahao2021bpublished,Zenil2023AIReviewarxivv3}.

If we disregard, for the sake of the argument, the potential difficulties described above and imagine that a given dataset provides a reliable source for scientific evidence, the question to be reflected upon would be about the possibility to mechanically discriminate patterns which might unveil causal relationships among variables of interest from spurious correlations \cite{Floridi2012BigDataepistemological,Smith2020,Kitchin2014,Calude2017}. A key criterion for the detection of causal relationships among variables is to discover the directionality of these variables, i.e., to uncover a causal chain (i.e., that rain causes the soil to be wet and not vice versa), or to show that two variables are the effect of the same cause (for instance, that thunder and lightning are caused by electric discharge). This reflects the Reichenbach principle: if two properties or kinds of events are probabilistically dependent, then they are causally connected in the sense that either one is a cause of the other (or vice versa), or both are effects of a common cause (where X is a cause of Y iff there leads a path of causal arrows from X to Y) \cite{Schurz2008Patternabduction,Glymour1991Causalinference}.

According to this criterion, the problem to be solved is how to discriminate if probabilistic dependent events are, in fact, causally related. 
This brings us back to the epistemological limitation discussed in Section~\ref{sectionDemons}.
Since Hume \cite{Hume2007edAnenquiry}, the problem of causality has haunted philosophers of science and epistemologists. Some philosophers, including Hume, go as far as suggest the we abandon the notion of causality, in logical and epistemological scenarios, by questioning the logical guarantee of reasoning about the future. If we cannot guarantee that the future will resemble the past, how can we establish, beyond psychological comfort, causal relationships among events?

Pearl \cite{Pearl2009Causalitybook} furnishes an interesting framework to reflect upon the possibility of attribution of causal relationship to variables of interest by proposing a methodological approach to causation based on three steps: 
\begin{description}[(i)]
\item[(i)]{observation by simple association between variables of interest does not guarantee the establishment of causality, only the discovery of correlational patterns;}

\item[(ii)]{observation and intervention of the variables of interest by, for instance, dividing a given set of elements into two groups to apply an intervention in one of these two groups and observe the differences and similarities between them (which could show which variables are causally relevant, or irrelevant, for a given set of events \cite{Bacon2011edNovumorganum,Mill2011edLogicRatiocinativeInductive}); and}

\item[(iii)]{imagination by means of counterfactual reasoning from which one might formulate hypothetical scenarios as a strategy to explain why an event is always followed by another event or to guess which common cause connects two or more events.}
\end{description}

This notion of an automated detection of relevant patterns by association among variables, algorithmic perturbations and/or counterfactual simulation seems to be a necessary paradigm shift in the scientific inquiry \cite{Zenil2023AIReviewarxivv3}. 

\section{Discussion}
\label{sectionConclusion}

We have introduced sufficient conditions for which the simplicity bubble effect (SBE) is a fundamental limitation of Big Data and machine learning methods that applies to formal knowledge production in lawlike universes.
In general, one cannot always avoid finding ``spuriously optimal models'', or ``spuriously true theories'', because eventually one cannot prevent a dataset to induce a learning algorithm or formal theory to find a local optimum that is deemed to be global, given that this deception is expected to occur anyway with non-negligible probability in randomly generated computably constructible datasets.
No formal-theoretic machine learning method can, in general, avoid the possibility of the very theory underlying this method being deceived into assigning itself more predictive powers than it actually has. 
The proofs are based on available data from which the optimal model is retrieved from in the first place (and upon which the predictions of the theory are corroborated) becoming sufficiently large in comparison to the complexity of the respective formal-theoretic system \cite{Abrahao2021dSBpaperarxiv}.

In the context of lawlike (computable) universes and formal learning systems, the epistemic consequences are more profound in the sense that these findings show that there is a ceiling above which formal knowledge (either produced by computation-assisted science or by artificial machines) actually cannot further decrease the probability of zemblanitous findings. 
It constitutes a double-edge-sword argument that may appear counter-intuitive at first glance. 
A formal theory that can only explain or predict (or is only corroborated by) a very small subset of events in a lawlike universe would be considered insufficiently valid or robust;
while a formal theory that explains too much can never neglect the possibility that it was deceived.

Furthermore, we argue this epistemological limitation can have impacts on societal dynamics.
Despite the advancements in technology of the past 100 years, two very different misconceptions on the state of technology are evident. 
The first, often dubbed ``post-digital'', is the belief that the digital revolution is over as a consequence of current practical limitations (such as processing speed and storage space), concluding that human intelligence is ultimately superior to machine intelligence. However, these limitations are likely temporary and far from the epistemic limits of computing. The other misconception is a total, unwavering belief that the ``promise'' of machine learning will result in a technology that achieves and surpasses human intelligence. 

Contrary to leaving the discussion and scientific research on the societal impacts of AI as a secondary problem in the everyday practice of AI developers~\cite{Birhane2022ValuesEncodedMachine}, we argue that a paradigmatic shift is necessary to understand these impacts: one where society looks deeper into the interplay between humans and machines, and how they are \textit{similar} and intertwined rather than how they are \textit{different}. 
We posit a `meta-digital' reexamination of these advancements that advocates for a more humanistic use of technology, but accepts that the digital revolution is still in its infancy and reflects the aforementioned paradigm shift. One in which AI is not necessarily viewed as inferior nor superior to human intelligence, but rather a self-reflexive view which illuminates such a limitation \textit{shared} by humans and computers: an epistemic limit of formal knowledge production itself to which any formal learning system, be it human or artificial, are prone.
This shared limitation in turn highlights the generality of the results presented in this article.
By confronting the property of digitally intermediated social dynamics being always prone to be trapped into zemblanitous findings, society will hopefully be compelled to think about, challenge, and reconsider the conceptual frameworks within which they understand and relate to technology. This, in turn, can lead to mechanisms that explain, predict, amplify, or mitigate certain collective digital-social dynamics. 

We believe that our society is at a crucial point in which governments and tech companies are collecting troves of personal data using unprecedented mass surveillance \cite{Zuboff2018surveillancecapitalism}. 
People can be surveilled without consent, or willingly contribute large amounts of data by subscribing to agreement terms that are required so they can use the respective tool, application, or platform.
In addition to the problem of the environmental impact of such technologies for the emissions of CO2 in the algorithm's training processes and respective technological waste, personal data is a commodity that can be used and sold. 
There is a concern regarding whether or not social networks have given rise to echo chambers and filter bubbles~\cite{Cinelli2021EchoChamberEffect,Terren2021EchoChambersSocial} that can polarize political opinions or influence digital behavior~\cite{Chitra2020AnalyzingImpactFilter,Kashima2021IdeologyCommunicationPolarization}. 
Because polarizing content engages users, algorithms could be created to sow and reinforce divisive beliefs, e.g. in order to maximize profits \cite{Lanier2018deletesocialmedia}.
Social networks and the underlying algorithms that intermediate social relationships can reflect political polarization and systemic inequality~\cite{Lanier2018deletesocialmedia,Noble2018algorithmsoppression}, and may also \textit{amplify} them~\cite{Cavassane2023BigDataSelffulfillingproph,ONeil2016weaponsmath}.
We argue these are societal impacts that highlight the importance of fostering the study of the theoretical or epistemological limitations discussed in this article.

Without a deeper understanding of the fundamental laws and nature of digitally intermediated societies, the very institutions created to protect people may become so irreconcilably affected that humanity as a whole is threatened. 
With the purpose of mitigating systemic inequalities and biases, we propose the fostering of scientific research in this field and the development of novel technological tools.
Those new tools should enable or reinforce that all members of society have more agency in the development and maintenance of a more humane and equitable political and economic framework.

\section{Acknowledgments}

Felipe S. Abrah\~{a}o acknowledges support from S\~{a}o Paulo Research Foundation (FAPESP), grants $2021$/$14501$-$8$ and $2023$/$05593$-$1$. Ricardo P. Cavassane acknowledges support from the National Council for Scientific and Technological Development (CNPq), Brazil, grant $150915$/$2024$-$1$, and from the Coordination for the Improvement of Higher Education Personnel (CAPES), Brazil, Finance Code $001$. Mariana Vitti-Rodrigues acknowledges the support from S\~{a}o Paulo Research Foundation (FAPESP), grant $2023$/$01408$-$5$.

\bibliographystyle{plainnat}
\bibliography{proceedings-v3-MBR-2023}

\begin{thebibliography}{55}
\providecommand{\natexlab}[1]{#1}
\providecommand{\url}[1]{\texttt{#1}}
\expandafter\ifx\csname urlstyle\endcsname\relax
  \providecommand{\doi}[1]{doi: #1}\else
  \providecommand{\doi}{doi: \begingroup \urlstyle{rm}\Url}\fi

\bibitem[Abrah{\~{a}}o et~al.(2020)Abrah{\~{a}}o, Wehmuth, and
  Ziviani]{Abrahao2020bsimplelocalrules}
Felipe~S. Abrah{\~{a}}o, Klaus Wehmuth, and Artur Ziviani.
\newblock {Emergence of complex data from simple local rules in a network
  game}.
\newblock In Edna~Alves de~Souza, Mariana~Claudia Broens, and Maria
  Eunice~Quilici Gonzalez, editors, \emph{Big Data: ethical and epistemological
  implications}, volume~89 of \emph{Cole\c{c}\~{a}o CLE}, pages 125--144.
  Cole\c{c}\~{a}o CLE e Editora FiloCzar, Campinas, 2020.
\newblock ISBN 978-65-87117-33-1.

\bibitem[Abrah{\~{a}}o et~al.(2023{\natexlab{a}})Abrah{\~{a}}o, Cavassane,
  Winter, and D'Ottaviano]{Abrahao2023SBZemblanityarxiv}
Felipe~S. Abrah{\~{a}}o, Ricardo~P. Cavassane, Michael Winter, and Itala M.~L.
  D'Ottaviano.
\newblock The simplicity bubble effect as a zemblanitous phenomenon in learning
  systems.
\newblock In \emph{Ninth Conference on Model-Based Reasoning, Abductive
  Cognition, Creativity}, Rome, 2023{\natexlab{a}}.
\newblock URL \url{https://www.mbr023rome.com/}.
\newblock Preprint available at: \url{https://arxiv.org/abs/2304.10681}.

\bibitem[Abrah{\~{a}}o et~al.(2023{\natexlab{b}})Abrah{\~{a}}o, Zenil, Porto,
  Winter, Wehmuth, and D'Ottaviano]{Abrahao2021dSBpaperarxiv}
Felipe~S. Abrah{\~{a}}o, Hector Zenil, Fabio Porto, Michael Winter, Klaus
  Wehmuth, and Itala M.~L. D'Ottaviano.
\newblock A simplicity bubble problem in formal-theoretic learning systems.
\newblock \emph{arXiv Preprints}, arXiv:2112.12275 [cs.IT], 2023{\natexlab{b}}.
\newblock URL \url{http://arxiv.org/abs/2112.12275v2}.

\bibitem[Abrahão and Arroyo(2023)]{Arroyo2023Nomicandsimplicity}
Felipe~S. Abrahão and Raoni Arroyo.
\newblock Nomic realism, simplicity, and the simplicity bubble effect.
\newblock \emph{arXiv}, arXiv:2310.17035 [physics.hist-ph], 2023.
\newblock URL \url{http://arxiv.org/abs/2310.17035}.

\bibitem[Abrahão and Zenil(2022)]{Abrahao2021bpublished}
Felipe~S. Abrahão and Hector Zenil.
\newblock Emergence and algorithmic information dynamics of systems and
  observers.
\newblock \emph{Philosophical Transactions of the Royal Society A:
  Mathematical, Physical and Engineering Sciences}, 380\penalty0 (2227), 2022.
\newblock ISSN 1364-503X.
\newblock \doi{10.1098/rsta.2020.0429}.

\bibitem[Alvarado(2023)]{Alvarado2023SimulatingScienceComputer}
Ram{\'o}n Alvarado.
\newblock \emph{Simulating {{Science}}: {{Computer Simulations}} as
  {{Scientific Instruments}}}, volume 479 of \emph{Synthese {{Library}}}.
\newblock Springer International Publishing, Cham, 2023.
\newblock ISBN 978-3-031-38646-6.
\newblock \doi{10.1007/978-3-031-38647-3}.

\bibitem[Andel(1994)]{Andel1994AnatomyUnsoughtFinding}
Pek~Van Andel.
\newblock Anatomy of the {{Unsought Finding}}. {{Serendipity}}: {{Orgin}},
  {{History}}, {{Domains}}, {{Traditions}}, {{Appearances}}, {{Patterns}} and
  {{Programmability}}.
\newblock \emph{The British Journal for the Philosophy of Science}, 45\penalty0
  (2):\penalty0 631--648, 1994.
\newblock ISSN 0007-0882.
\newblock \doi{10.1093/bjps/45.2.631}.

\bibitem[Anderson(2008)]{Anderson2008Endoftheory}
Chris Anderson.
\newblock The end of theory: The data deluge makes the scientific method
  obsolete.
\newblock \emph{Wired magazine}, 16\penalty0 (7):\penalty0 16--07, 2008.

\bibitem[Anishchenko et~al.(2021)Anishchenko, Pellock, Chidyausiku, Ramelot,
  Ovchinnikov, Hao, Bafna, Norn, Kang, Bera, DiMaio, Carter, Chow, Montelione,
  and Baker]{Anishchenko2021denovodesigndeeplearning}
Ivan Anishchenko, Samuel~J. Pellock, Tamuka~M. Chidyausiku, Theresa~A. Ramelot,
  Sergey Ovchinnikov, Jingzhou Hao, Khushboo Bafna, Christoffer Norn, Alex
  Kang, Asim~K. Bera, Frank DiMaio, Lauren Carter, Cameron~M. Chow, Gaetano~T.
  Montelione, and David Baker.
\newblock De novo protein design by deep network hallucination.
\newblock \emph{Nature}, 600\penalty0 (7889):\penalty0 547--552, 2021.
\newblock ISSN 0028-0836, 1476-4687.
\newblock \doi{10.1038/s41586-021-04184-w}.

\bibitem[Bacon(2011)]{Bacon2011edNovumorganum}
Francis Bacon.
\newblock {NOVUM} {ORGANUM}.
\newblock In James Spedding, Robert~Leslie Ellis, and Douglas~Denon Heath,
  editors, \emph{The {Works} of {Francis} {Bacon}}. Cambridge University Press,
  1 edition, 2011.
\newblock ISBN 978-1-108-04064-8 978-1-139-14954-9.
\newblock \doi{10.1017/CBO9781139149549.019}.

\bibitem[Birhane et~al.(2022)Birhane, Kalluri, Card, Agnew, Dotan, and
  Bao]{Birhane2022ValuesEncodedMachine}
Abeba Birhane, Pratyusha Kalluri, Dallas Card, William Agnew, Ravit Dotan, and
  Michelle Bao.
\newblock The {{Values Encoded}} in {{Machine Learning Research}}.
\newblock In \emph{2022 {{ACM Conference}} on {{Fairness}}, {{Accountability}},
  and {{Transparency}}}, pages 173--184. ACM, 2022.
\newblock ISBN 978-1-4503-9352-2.
\newblock \doi{10.1145/3531146.3533083}.

\bibitem[Boyd(1998)]{Boyd1998}
William Boyd.
\newblock \emph{Armadillo}.
\newblock Hamish Hamilton, 1998.

\bibitem[Callaway(2022)]{Callaway2022AlphaFoldAI}
Ewen Callaway.
\newblock What's next for {AlphaFold} and the {AI} protein-folding revolution.
\newblock \emph{Nature}, 604\penalty0 (7905):\penalty0 234--238, 2022.
\newblock ISSN 0028-0836, 1476-4687.
\newblock \doi{10.1038/d41586-022-00997-5}.

\bibitem[Calude and Longo(2017)]{Calude2017}
Cristian~S. Calude and Giuseppe Longo.
\newblock The {Deluge} of {Spurious} {Correlations} in {Big} {Data}.
\newblock \emph{Foundations of Science}, 22\penalty0 (3):\penalty0 595--612,
  September 2017.
\newblock ISSN 1233-1821, 1572-8471.
\newblock \doi{10.1007/s10699-016-9489-4}.
\newblock URL \url{http://link.springer.com/10.1007/s10699-016-9489-4}.

\bibitem[Cavassane(2022)]{Cavassane2022}
Ricardo~P. Cavassane.
\newblock Zemblanity and big {Data}: the ugly truths the algorithms remind us
  of.
\newblock \emph{Acta Scientiarum. Human and Social Sciences}, 44\penalty0 (1),
  2022.
\newblock ISSN 1807-8656, 1679-7361.
\newblock \doi{10.4025/actascihumansoc.v44i1.62246}.
\newblock URL
  \url{https://periodicos.uem.br/ojs/index.php/ActaSciHumanSocSci/article/view/62246}.

\bibitem[Cavassane et~al.(2023)Cavassane, Abrah{\~{a}}o, and
  D'Ottaviano]{Cavassane2023BigDataSelffulfillingproph}
Ricardo~P. Cavassane, Felipe~S. Abrah{\~{a}}o, and Itala M.~L. D'Ottaviano.
\newblock Big {Data} and the {Emergence} of {Zemblanity} and
  {Self}-{Fulfilling} {Prophecies}.
\newblock In \emph{{Cognition} \& {Modeling}, {Proceedings} of 11th
  {International} {Meeting} on {Informational}, {Knowledge} and {Action}},
  {Cognition} \& {Modeling}, 2023.
\newblock URL \url{https://philpapers.org/rec/CAVBDA}.

\bibitem[Chaitin(2012)]{Chaitin2012}
Gregory Chaitin.
\newblock \emph{A Computable Universe}, chapter Life as Evolving Software.
\newblock World Scientific, 2012.

\bibitem[Chitra and Musco(2020)]{Chitra2020AnalyzingImpactFilter}
Uthsav Chitra and Christopher Musco.
\newblock Analyzing the {{Impact}} of {{Filter Bubbles}} on {{Social Network
  Polarization}}.
\newblock In \emph{Proceedings of the 13th {{International Conference}} on
  {{Web Search}} and {{Data Mining}}}, pages 115--123. ACM, 2020.
\newblock ISBN 978-1-4503-6822-3.
\newblock \doi{10.1145/3336191.3371825}.

\bibitem[Cinelli et~al.(2021)Cinelli, De~Francisci~Morales, Galeazzi,
  Quattrociocchi, and Starnini]{Cinelli2021EchoChamberEffect}
Matteo Cinelli, Gianmarco De~Francisci~Morales, Alessandro Galeazzi, Walter
  Quattrociocchi, and Michele Starnini.
\newblock The echo chamber effect on social media.
\newblock \emph{Proceedings of the National Academy of Sciences}, 118\penalty0
  (9), 2021.
\newblock ISSN 0027-8424.
\newblock \doi{10.1073/pnas.2023301118}.

\bibitem[Collmann et~al.(2016)Collmann, FitzGerald, Wu, Kupersmith, and
  Matei]{Collmann2016EthicalBigDataaouthumans}
Jeff Collmann, Kevin~T. FitzGerald, Samantha Wu, Joel Kupersmith, and
  Sorin~Adam Matei.
\newblock Data {Management} {Plans}, {Institutional} {Review} {Boards}, and the
  {Ethical} {Management} of {Big} {Data} {About} {Human} {Subjects}.
\newblock In Jeff Collmann and Sorin~Adam Matei, editors, \emph{Ethical
  {Reasoning} in {Big} {Data}}, pages 141--184. Springer International
  Publishing, Cham, 2016.
\newblock ISBN 978-3-319-28420-0.
\newblock \doi{10.1007/978-3-319-28422-4{\_}10}.

\bibitem[Copeland(2019)]{Copeland2019}
Samantha Copeland.
\newblock On serendipity in science: discovery at the intersection of chance
  and wisdom.
\newblock \emph{Synthese}, 196\penalty0 (6), 2019.
\newblock ISSN 1573-0964.
\newblock \doi{10.1007/s11229-017-1544-3}.

\bibitem[Floridi(2012)]{Floridi2012BigDataepistemological}
Luciano Floridi.
\newblock Big {Data} and {Their} {Epistemological} {Challenge}.
\newblock \emph{Philosophy \& Technology}, 25\penalty0 (4):\penalty0 435--437,
  2012.
\newblock ISSN 2210-5433, 2210-5441.
\newblock \doi{10.1007/s13347-012-0093-4}.

\bibitem[Glymour et~al.(1991)Glymour, Spirtes, and
  Scheines]{Glymour1991Causalinference}
C.~Glymour, P.~Spirtes, and R.~Scheines.
\newblock Causal inference.
\newblock \emph{Erkenntnis}, 35\penalty0 (1-3):\penalty0 151--189, 1991.
\newblock ISSN 0165-0106, 1572-8420.
\newblock \doi{10.1007/BF00388284}.

\bibitem[Goodfellow et~al.(2016)Goodfellow, Bengio, and
  Courville]{Goodfellow2016}
Ian Goodfellow, Yoshua Bengio, and Aaron Courville.
\newblock \emph{Deep Learning}.
\newblock MIT Press, 2016.
\newblock \url{http://www.deeplearningbook.org}.

\bibitem[Hernández-Orozco et~al.(2021)Hernández-Orozco, Zenil, Riedel,
  Uccello, Kiani, and Tegnér]{HernandezOrozco2021}
Santiago Hernández-Orozco, Hector Zenil, Jürgen Riedel, Adam Uccello,
  Narsis~A. Kiani, and Jesper Tegnér.
\newblock Algorithmic {Probability}-{Guided} {Machine} {Learning} on
  {Non}-{Differentiable} {Spaces}.
\newblock \emph{Frontiers in Artificial Intelligence}, 3:\penalty0 567356,
  January 2021.
\newblock ISSN 2624-8212.
\newblock \doi{10.3389/frai.2020.567356}.
\newblock URL
  \url{https://www.frontiersin.org/articles/10.3389/frai.2020.567356/full}.

\bibitem[Hume and Buckle(2007)]{Hume2007edAnenquiry}
David Hume and Stephen Buckle.
\newblock \emph{An enquiry concerning human understanding and other writings}.
\newblock Cambridge texts in the history of philosophy. Cambridge University
  Press, Cambridge ; New York, 2007.
\newblock ISBN 978-0-521-84340-9.

\bibitem[Kashima et~al.(2021)Kashima, Perfors, Ferdinand, and
  Pattenden]{Kashima2021IdeologyCommunicationPolarization}
Yoshihisa Kashima, Andrew Perfors, Vanessa Ferdinand, and Elle Pattenden.
\newblock Ideology, communication and polarization.
\newblock \emph{Philosophical Transactions of the Royal Society B: Biological
  Sciences}, 376\penalty0 (1822), 2021.
\newblock ISSN 0962-8436,.
\newblock \doi{10.1098/rstb.2020.0133}.

\bibitem[Kitchin(2014)]{Kitchin2014}
Rob Kitchin.
\newblock Big {Data}, new epistemologies and paradigm shifts.
\newblock \emph{Big Data \& Society}, 1\penalty0 (1), 2014.
\newblock ISSN 2053-9517.
\newblock \doi{10.1177/2053951714528481}.

\bibitem[Lanier(2018)]{Lanier2018deletesocialmedia}
Jaron Lanier.
\newblock \emph{{Ten Arguments for Deleting Your Social Media Accounts Right
  Now}}.
\newblock Henry Holt and Company, 2018.
\newblock ISBN 9781250196699.

\bibitem[Leonelli(2014)]{Leonelli2014QuantityepistemologyBigData}
S~Leonelli.
\newblock What difference does quantity make? {On} the epistemology of {Big}
  {Data} in biology.
\newblock \emph{Big Data \& Society}, 1\penalty0 (1), 2014.
\newblock ISSN 2053-9517.
\newblock \doi{10.1177/2053951714534395}.

\bibitem[Leonelli(2023)]{Leonelli2023OpenScience}
Sabina Leonelli.
\newblock \emph{Philosophy of {Open} {Science}}.
\newblock Cambridge University Press, 1 edition, 2023.
\newblock ISBN 978-1-00-941636-8.
\newblock \doi{10.1017/9781009416368}.

\bibitem[Mayer-Schönberger and Cukier(2013)]{Cukier2013}
Viktor Mayer-Schönberger and Kenneth Cukier.
\newblock \emph{Big data: a revolution that will transform how we live, work
  and think}.
\newblock Houghton Mifflin Harcourt, 2013.

\bibitem[Merton(1948)]{Merton1948}
Robert~K. Merton.
\newblock The bearing of empirical research upon the development of social
  theory.
\newblock \emph{American Sociological Review}, 13\penalty0 (5), 1948.
\newblock ISSN 0003-1224, 1939-8271.
\newblock \doi{10.2307/2087142}.
\newblock URL \url{https://www.jstor.org/stable/2087142}.

\bibitem[Merton(2000)]{Merton2000SocialTheorySocial}
Robert~King Merton.
\newblock \emph{Social Theory and Social Structure}.
\newblock Free Press, New York, NY, enlarged ed., [nachdr.] edition, 2000.
\newblock ISBN 978-0-02-921130-4.

\bibitem[Mill(2011)]{Mill2011edLogicRatiocinativeInductive}
John~Stuart Mill.
\newblock \emph{A {System} of {Logic}, {Ratiocinative} and {Inductive}: {Being}
  a {Connected} {View} of the {Principles} of {Evidence}, and the {Methods} of
  {Scientific} {Investigation}}.
\newblock Cambridge University Press, 1 edition, 2011.
\newblock ISBN 978-1-108-04088-4.
\newblock \doi{10.1017/CBO9781139149839}.

\bibitem[Noble(2018)]{Noble2018algorithmsoppression}
Safiya~Umoja Noble.
\newblock \emph{{Algorithms of Oppression: How Search Engines Reinforce
  Racism}}.
\newblock NYU Press, 2018.
\newblock ISBN 9781479866762.

\bibitem[O'Neil(2016)]{ONeil2016weaponsmath}
Cathy O'Neil.
\newblock \emph{Weapons of Math Destruction: {{How}} Big Data Increases
  Inequality and Threatens Democracy}.
\newblock Crown, 2016.
\newblock ISBN 0-553-41881-5.

\bibitem[Pearl(2009)]{Pearl2009Causalitybook}
Judea Pearl.
\newblock \emph{Causality: Models, Reasoning and Inference}.
\newblock Cambridge University Press, USA, 2nd edition, 2009.
\newblock ISBN 052189560X.

\bibitem[Scholkopf et~al.(2021)Scholkopf, Locatello, Bauer, Ke, Kalchbrenner,
  Goyal, and Bengio]{Scholkopf2021}
Bernhard Scholkopf, Francesco Locatello, Stefan Bauer, Nan~Rosemary Ke, Nal
  Kalchbrenner, Anirudh Goyal, and Yoshua Bengio.
\newblock Toward {Causal} {Representation} {Learning}.
\newblock \emph{Proceedings of the IEEE}, 109\penalty0 (5):\penalty0 612--634,
  May 2021.
\newblock ISSN 0018-9219, 1558-2256.
\newblock \doi{10.1109/JPROC.2021.3058954}.
\newblock URL \url{https://ieeexplore.ieee.org/document/9363924/}.

\bibitem[Schurz(2008)]{Schurz2008Patternabduction}
G.~Schurz.
\newblock Patterns of abduction.
\newblock \emph{Synthese}, 164\penalty0 (2):\penalty0 201--234, 2008.
\newblock ISSN 0039-7857, 1573-0964.
\newblock \doi{10.1007/s11229-007-9223-4}.

\bibitem[Smith(2020)]{Smith2020}
Gary Smith.
\newblock The paradox of big data.
\newblock \emph{SN Applied Sciences}, 2\penalty0 (6):\penalty0 1041, 2020.
\newblock ISSN 2523-3963, 2523-3971.
\newblock \doi{10.1007/s42452-020-2862-5}.
\newblock URL \url{http://link.springer.com/10.1007/s42452-020-2862-5}.

\bibitem[Terren et~al.(2021)Terren, {Open University of Catalonia},
  {Borge-Bravo}, and {Open University of
  Catalonia}]{Terren2021EchoChambersSocial}
Ludovic Terren, {Open University of Catalonia}, Rosa {Borge-Bravo}, and {Open
  University of Catalonia}.
\newblock Echo {{Chambers}} on {{Social Media}}: {{A Systematic Review}} of the
  {{Literature}}.
\newblock \emph{Review of Communication Research}, 9:\penalty0 99--118, 2021.
\newblock ISSN 22554165.
\newblock \doi{10.12840/ISSN.2255-4165.028}.

\bibitem[Uthamacumaran et~al.(2024)Uthamacumaran, Abrah{\~a}o, Kiani, and
  Zenil]{Uthamacumaran2023arxivMSpaper}
Abicumaran Uthamacumaran, Felipe~S. Abrah{\~a}o, Narsis~A. Kiani, and Hector
  Zenil.
\newblock On the salient limitations of the methods of assembly theory and
  their classification of molecular biosignatures.
\newblock \emph{npj Systems Biology and Applications}, 10\penalty0
  (1):\penalty0 82, August 2024.
\newblock ISSN 2056-7189.
\newblock \doi{10.1038/s41540-024-00403-y}.

\bibitem[Walpole()]{WalpoleCorrespondence}
Horace Walpole.
\newblock Horace walpole’s correspondence: Yale edition.
\newblock URL \url{https://libsvcs-1.its.yale.edu/hwcorrespondence/}.

\bibitem[Winter(2020)]{Winter2020}
Michael Winter.
\newblock Meta+phenomenology: {Primer} {Towards} a {Phenomenology} {Formally}
  {Based} on {Algorithmic} {Information} {Theory} and {Metabiology}.
\newblock In \emph{Unravelling {Complexity}}, pages 317--334. World Scientific,
  feb 2020.

\bibitem[Witten et~al.(2017)Witten, Frank, Hall, and Pal]{Witten2017}
Ian~H. Witten, Eibe Frank, Mark~A. Hall, and Christopher~J. Pal, editors.
\newblock \emph{Data Mining}.
\newblock Morgan Kaufmann, fourth edition edition, 2017.
\newblock ISBN 978-0-12-804291-5.
\newblock \doi{10.1016/C2015-0-02071-8}.
\newblock URL
  \url{https://www.sciencedirect.com/science/article/pii/B9780128042915000052}.

\bibitem[Wolpert and Macready(1997)]{Wolpert1997}
D.H. Wolpert and W.G. Macready.
\newblock No free lunch theorems for optimization.
\newblock \emph{IEEE Transactions on Evolutionary Computation}, 1\penalty0
  (1):\penalty0 67--82, 1997.
\newblock \doi{10.1109/4235.585893}.

\bibitem[Zenil(2013)]{Zenil2013}
Hector Zenil, editor.
\newblock \emph{{A computable universe: understanding and exploring nature as
  computation}}.
\newblock World Scientific Publishing, 2013.
\newblock ISBN 978-981-4447-78-2.

\bibitem[Zenil et~al.(2018)Zenil, Hern{\'{a}}ndez-Orozco, Kiani, Soler-Toscano,
  Rueda-Toicen, and Tegn{\'{e}}r]{Zenil2018}
Hector Zenil, Santiago Hern{\'{a}}ndez-Orozco, Narsis Kiani, Fernando
  Soler-Toscano, Antonio Rueda-Toicen, and Jesper Tegn{\'{e}}r.
\newblock {A Decomposition Method for Global Evaluation of Shannon Entropy and
  Local Estimations of Algorithmic Complexity}.
\newblock \emph{Entropy}, 20\penalty0 (8):\penalty0 605, aug 2018.
\newblock ISSN 1099-4300.
\newblock \doi{10.3390/e20080605}.
\newblock URL \url{http://www.mdpi.com/1099-4300/20/8/605}.

\bibitem[Zenil et~al.(2019{\natexlab{a}})Zenil, Kiani, Marabita, Deng, Elias,
  Schmidt, Ball, and Tegn{\'{e}}r]{Zenil2019dAIDcalculusiScience}
Hector Zenil, Narsis~A. Kiani, Francesco Marabita, Yue Deng, Szabolcs Elias,
  Angelika Schmidt, Gordon Ball, and Jesper Tegn{\'{e}}r.
\newblock {An Algorithmic Information Calculus for Causal Discovery and
  Reprogramming Systems}.
\newblock \emph{iScience}, 19:\penalty0 1160--1172, sep 2019{\natexlab{a}}.
\newblock \doi{10.1016/j.isci.2019.07.043}.

\bibitem[Zenil et~al.(2019{\natexlab{b}})Zenil, Kiani, Zea, and
  Tegn{\'{e}}r]{Zenil2019}
Hector Zenil, Narsis~A. Kiani, Allan~A. Zea, and Jesper Tegn{\'{e}}r.
\newblock {Causal deconvolution by algorithmic generative models}.
\newblock \emph{Nature Machine Intelligence}, 1\penalty0 (1):\penalty0 58--66,
  jan 2019{\natexlab{b}}.
\newblock ISSN 2522-5839.
\newblock \doi{10.1038/s42256-018-0005-0}.
\newblock URL \url{http://www.nature.com/articles/s42256-018-0005-0}.

\bibitem[Zenil et~al.(2020)Zenil, Kiani, Abrah{\~{a}}o, and
  Tegn{\'{e}}r]{Zenil2020cAIDscholarpedia}
Hector Zenil, Narsis Kiani, Felipe Abrah{\~{a}}o, and Jesper Tegn{\'{e}}r.
\newblock {Algorithmic Information Dynamics}.
\newblock \emph{Scholarpedia Journal}, 15\penalty0 (7):\penalty0 53143, 2020.
\newblock ISSN 1941-6016.
\newblock \doi{10.4249/scholarpedia.53143}.

\bibitem[Zenil et~al.(2023{\natexlab{a}})Zenil, Adams, and
  Abrah{\~{a}}o]{Zenil2023arxivETpaper}
Hector Zenil, Alyssa Adams, and Felipe~S. Abrah{\~{a}}o.
\newblock Optimal spatial deconvolution and message reconstruction from a large
  generative model of models.
\newblock \emph{arXiv Preprints}, arXiv:1802.05843 [cs.DS], 2023{\natexlab{a}}.
\newblock \doi{10.48550/arXiv.1802.05843}.
\newblock URL \url{https://arxiv.org/abs/1802.05843}.

\bibitem[Zenil et~al.(2023{\natexlab{b}})Zenil, Tegnér, Abrahão, Lavin,
  Kumar, Frey, Weller, Soldatova, Bundy, Jennings, Takahashi, Hunter, Dzeroski,
  Briggs, Gregory, Gomes, Rowe, Evans, Kitano, and
  King]{Zenil2023AIReviewarxivv3}
Hector Zenil, Jesper Tegnér, Felipe~S. Abrahão, Alexander Lavin, Vipin Kumar,
  Jeremy~G. Frey, Adrian Weller, Larisa Soldatova, Alan~R. Bundy, Nicholas~R.
  Jennings, Koichi Takahashi, Lawrence Hunter, Saso Dzeroski, Andrew Briggs,
  Frederick~D. Gregory, Carla~P. Gomes, Jon Rowe, James Evans, Hiroaki Kitano,
  and Ross King.
\newblock The {Future} of {Fundamental} {Science} {Led} by {Generative}
  {Closed}-{Loop} {Artificial} {Intelligence}.
\newblock arXiv:2307.07522 [cs], 2023{\natexlab{b}}.
\newblock URL \url{http://arxiv.org/abs/2307.07522}.

\bibitem[Zuboff(2018)]{Zuboff2018surveillancecapitalism}
Shoshana Zuboff.
\newblock \emph{{The Age of Surveillance Capitalism: The Fight for a Human
  Future at the New Frontier of Power}}.
\newblock PublicAffairs, 2018.
\newblock ISBN 1610395697.

\end{thebibliography}
\end{document}